% be compiled by TEX (not latex)
%Version September.15.2000
\magnification=1200 \baselineskip=13pt \hsize=16.5 true cm \vsize=20 true cm
\def\parG{\vskip 10pt} \font\bbold=cmbx10 scaled\magstep2
\centerline{{\sl Trends in Statistical Physics} (2001)}\parG

\centerline{\bbold Dynamic Drop Models}\parG

\centerline{P.M.C. de Oliveira$^1$, T.J.P. Penna$^1$, A.R. Lima$^2$,}\par
\centerline{J.S. S\'a Martins$^3$, C. Moukarzel$^{1,4}$ and C.A.F.
Leite$^1$}\parG
\centerline{$^1$ Instituto de F\'\i sica, Universidade Federal
Fluminense}\par
\centerline{av. Litor\^anea s/n, Boa Viagem, Niter\'oi RJ, Brazil 24210-340}\par
\centerline{$^2$ P.M.M.H., \'Ecole Sup\'erieure de Physique et de Chimie
Industrielles (ESPCI)}\par
\centerline{10, rue Vauquelin, 75231 Paris Cedex 05, France}\par
\centerline{$^3$ Colorado Center for Chaos and Complexity, CIRES, and}\par
\centerline{Department of  Physics, University of Colorado, Boulder, CO,
USA, 80309.}\par
\centerline{$^4$ Departamento de Fisica Aplicada, CINVESTAV Unidad Merida}\par
\centerline{Antigua carretera a Progreso, km. 6, 97310 Merida, Yuc.,
Mexico}\par
\centerline{e-mail: PMCO @ IF.UFF.BR}

\vskip 0.1cm\leftskip=0.8cm\rightskip=0.8cm 

{\bf Abstract}\parG

	We follow the dynamic evolution of a cluster of Ising spins
pointing $up$ surrounded by other spins pointing $down$, on a lattice. The
cluster represents a liquid drop. Under a microscopic point of view, the
short range ferromagnetic coupling between these spins plays the role of
the van der Waals attraction. Alternatively, under a macroscopic point of
view, the same ferromagnetic coupling gives rise to the surface tension
along the drop boundary. This naive model is applied to the study of
different systems, out of thermodynamic equilibrium. For each such a
system, other interaction terms can be included, for instance an external
magnetic field with a downwards uniform gradient, representing Earth's
gravity. Also, for each system, proper dynamic rules and boundary
conditions are adopted.

	The behaviour of such a drop is monitored as a function of time,
through computer simulations. Many quantities of interest, in particular
those related to drop fragmentation, were measured and the results were
compared with available experimental data. The real systems we have in
mind are exemplified by water drops falling from a leaky faucet, nuclear
multifragmentation, mercury drops falling on the ground, magnetic
hysteresis curves, and interface roughness.\par

\leftskip=0pt\rightskip=0pt\parG


\vfill\eject

{\bf I -- Introduction}\parG

	The stationary geometry of a water drop pending from a vertical
wall is an old mathematical puzzle (see [1] and references therein). A
stochastic model [1] was applied as an alternative way to solve this
puzzle. The drop is represented by a cluster of Ising spins pointing $up$
($S_i = +1$) on a square lattice (only the projection of the drop on the
vertical plane perpendicular to the wall is of interest). All other
lattice spins point $down$ ($S_i = -1$), representing the air around the
water drop. These spins are coupled through a first-and-second neighbour
ferromagnetic interaction term strong enough to avoid the drop
fragmentation. A non-uniform external magnetic field pointing also $up$ is
included, with the same intensity along each horizontal lattice row.
However, this intensity increases uniformly when one goes from each row to
the next one just below, i.e. there is a uniform vertical downward
gradient which mimics the Earth's gravity. The Hamiltonian reads

$$ {\cal H} = - \sum_{<ij>} S_i S_j - g \sum_r r \sum_i S_i\,\,\,\, ,
\eqno(1)$$

\noindent where the first sum runs over all pairs $<ij>$ of spins which
are first or second neighbours on the square lattice. The index $r = 0, 1,
2 \dots$ counts the lattice rows downwards, and $g$ is a free parameter
representing the Earth's gravity intensity. The terms $up$ and $down$
refer to the Ising spin orientation only, they have nothing to do with the
vertical axis of the square lattice where the drop is represented: they
could be replaced by $wet$ and $dry$, respectively.

	The drop starts as half a circle glued on the upper part of the
vertical wall represented by an external boundary lattice column, as
schematically shown in figure I.1. Along this column, the upper spins
point $up$, i.e. those in front of the drop diameter, while all other
spins below it point $down$: the upper part of the wall is $wet$, in
contact with the drop, but it remains $dry$ below the drop. The spins
along this boundary lattice column are kept {\bf fixed} in this situation
during all the drop time evolution. The ferromagnetic interaction term
between the lattice spins and these fixed boundary spins keeps the drop
glued on the wall and avoids it to slide downwards. A Monte Carlo updating
of the spins outside this boundary wall is then performed in order to
achieve equilibrium. Eventually, the geometrical aspect of the drop
reaches statistical equilibrium; its {\bf average} aspect can then be
measured and compared with real drops [1]. Each updating step consists in
choosing a random pair of spins, one pointing $up$ along the drop's
current inner boundary, the other pointing $down$ along the current outer
boundary. Then, both are flipped if this movement decreases the total
energy given by equation (1) plus the wall boundary condition.
Accordingly, the drop mass (the total number of spins pointing $up$)
remains unchanged, and only its geometrical shape is equilibrated by this
relaxation process.

	A very small artificial temperature could be included in order to
decide to flip or not each previously tossed pair of spins. However, the
only role played by this further artifact would be to avoid some possible
trappings into metastable situations, during the evolution. According to
our numerical experiments, the randomness in tossing the pair of spins to
be flipped is enough to equilibrate the system, without need to resort to
any further artificial thermal noise. Also according to our numerical
experiments, one does not need to include second neighbour coupling into
equation (1). The usual first-neighbour only interaction is enough.
However, in this case, the geometrical shape of the drop is not so smooth,
presenting mostly vertical or horizontal pieces along the surface: either
a larger lattice or a larger number of averaging samples would be required
in order to obtain the same accuracy.

	The ferromagnetic coupling represented by the first summation of
equation (1) obviously mimics an attraction between neighbouring water
``molecules'', i.e. a ``van der Waals term'' keeping the drop connected
into a single macroscopic piece. The same term can be rewritten as $2\sum
(1-\delta_{S_i,S_j})$ plus an additive unimportant constant. Under this
point of view, only the drop boundary contributes to the energy, i.e. it
represents a surface tension which tends to minimize the drop perimeter.
Indeed, only at the drop boundary one can find two neighbouring spins
$S_i$ and $S_j$, one pointing $up$ inside the drop, the other $down$
outside, contributing to the above sum. Otherwise, far from the boundary,
one has $S_i = S_j$ (both neighbouring spins pointing $up$ inside the
drop, or both $down$ outside), with no contribution to the above sum.
Thus, the time evolution minimizing the energy given by the complete
Hamiltonian (1) represents a competitive process between the first
summation which tends to keep the drop spherical, and the second summation
which tends to stretch it downwards. The Monte Carlo simulation performed
in [1] studies the equilibrium compromise between these two opposed
trends. In this case, the aim is simply to measure the geometrical shape
of the drop at equilibrium, after performing some transient relaxation
movements. Indeed, the geometry obtained in this way compares very well
with real drops [1].

	The same model can also be applied to the study of other systems
which are really out of equilibrium. This was done by the authors during
the last 8 years [2-12]. In this case, the interest is not to measure the
equilibrium properties, but to follow the dynamic evolution of the system
during its path towards equilibrium. This is the case of
multifragmentation in both nuclear matter (section III) and mercury drops
falling on the ground (section IV). It is also the case of annealing
multilayered interfaces where a moving domain wall replaces the drop
(section VI). Other systems are continuously fed from outside, and never
reach equilibrium. Examples are water drops falling from a leaky faucet
(section II) or magnetic hysteresis loops (section V). In all cases, the
same basic lattice drop model described above is adopted, with the surface
tension term. In each case, some further ingredients are included into
this basic model, according to the particular features of the system under
study. The present text is a review of these works.

\vskip 1cm
{\bf II -- Leaky Faucets}\parG

	The sequence of drops falling from a leaky faucet is a nice
example of complex, long-term memory dynamics working in a simple real
system. It became a paradigm for such dynamics after the work of R. Shaw
[13]. Since then, many experiments were performed [14-28], running from
the very simple to the very sophisticated. A spring-mass like model, first
proposed by Shaw [13, 18, 29, 30] and improved by Fuchikami, Kyono and
Ishioka [31,32], also provide very interesting descriptions of the
complexity observed during drop formation. Apart from the geometrical
aspect of a single drop itself, for which there are very beautiful
experiments and simulations (see [5,19,23,28,31,32] and references
therein), another interesting object to study is the sequence of time
intervals measured between successive drops. This time series displays the
quoted long-term memory dynamics. Very long time series, with over a
million successive drops, taken with extreme care are available
[20,4,21,23-26]. In spite of the term $chaotic$ in the title of the
original study [13], nobody knows if this dynamics is really chaotic:
according to the available data, it seems to be complex, not chaotic. In
[4], for instance, long-term anticorrelations were measured with very good
accuracy, displaying the value $\alpha = 0.00$ for the Hurst exponent (see
below) for many distinct experimental conditions. The same value with the
same accuracy was measured from computer simulations [4] also for many
distinct simulated conditions. Anyway, a conclusive definition of the
character (chaotic, regular or complex) of this dynamics can be obtained
only by measuring its Lyapunov exponent $\lambda$ ($> 0$, $< 0$ or $= 0$,
respectively). Currently available data give $\lambda \approx 0$.

	In this section we review some results obtained with a lattice
drop model [2-5] based on that proposed by Manna, Herrmann and Landau [1]
presented in the introduction. The model is very simple compared to others
designed to simulate the formation of a single drop. However, as we have
shown during the last years [2-5] it is also able to reproduce dynamical
and morphological aspects of real drops.

	In the model, the faucet is represented by $W$ elements of fluid
fixed in the center of the first, top lattice row from which the drop
pends. The flux of water is simulated by introducing $F$ elements directly
{\bf on the outer boundary} of the drop at fixed intervals of time.
Between sucessive injections of fluid, $N \times F$ relaxations are
performed. This new parameter $N$ is related with the viscosity of the
fluid. The time can be defined as proportional to the number of
relaxations performed since the beginning of the simulation. The
disconnection of the drop is verified after each relaxation, and if it
occurs, the disconnected piece is removed. The remaining fluid still
pending from the faucet sets the initial condition for the next drop: it
is responsible for the long-term memory.

	The first question one can formulate about this model is if it
could describe the morphology of real drops [5]. In order to answer this
question we can follow some quantities like the volume, perimeter and
center of mass height (measured downwards from the faucet) to compare with
data available from real drops [23]. In figure II.1 we show some steps of
the drop formation for $W=40$, $N=150$, $g=0.1$ and $F=20$. The numbers in
the figure are the times when each image was obtained, counted since the
last drop disconnection. We did not perform any averaging process to
smooth the boundary.

	In figures II.2, 3 and 4 we present the center of mass height
$versus$ time, area $versus$ perimeter and perimeter/area $versus$ time
for the same drop in figure II.1. These results are essentially equivalent
to that obtained from real drops [23]. Some features, like the formation
of the ``neck'' before the complete disconnection of the drop, are
present. A feature which is not present in the simulational results are
the oscillations of the drop [23,31,32]. This is expected once the model
does not consider inertia.

	Our results show that the morphology of real drops can be
reproduced by this simple model. We now turn our attention to a dynamics
characterisation, through the analysis of the sequences of time intervals
measured between successive drops, both in simulations and the experiment.
This was first done by de Oliveira and Penna [2,3] and using more
sophisticated techniques in [4]. Let's call $\{B(n)\}$ the series of time
intervals between successive drops. The main tool used to study such
series is the reconstruction of the attractor of the system using return
maps. The first return map, for instance, is obtained by plotting $B(n+1)$
$versus$ $B(n)$. In figure II.5 we show some first return maps constructed
from time series obtained with the lattice model described in this
chapter. These attractors and many others reproduce very well the real
ones, mainly that produced by tiny faucets [17]. Different attractors are
obtained varying the flux, just as in the experimental procedure.

	The time series measured from successive drops falling from leaky
faucets, experimental and simulated, are very similar [4] to the ones
obtained from healthy heart beat intervals [33]. The authors of these
references used the mean fluctuation in order to analyze the time series.
This quantity is defined as (see, for instance [34])

$$F(n) =\,\, <|{B(n'+n) - B(n')}|>_{n'} \,\,\,\, , \eqno(2)$$

\noindent where the brackets correspond to the average over all $n'$
values. It measures the correlations at each time scale $n$, and behaves
like

$$F(n) \sim n^{\alpha} \,\,\,\, , \eqno(3)$$

\noindent in the long-term limit $n \to \infty$. For healthy heart beat
series the Hurst exponent is $\alpha \cong 0$. This means that the time
series are in the extreme case of anticorrelation. For a random walk, the
Hurst exponent is exactly $\alpha = 1/2$ (no long-range correlations at
all), whereas $\alpha = 1$ for completely correlated systems. Actually, if
the elements $B(n)$ of the time series are randomly shuffled, the exponent
changes to $\alpha = 1/2$, which shows that we are dealing with a
phenomenon dictated by the {\bf chronology} according to which the
successive heart beats occur. As obtained in [4], for both experimental
and simulated time series, exactly the same thing occurs for the drops
falling from leaky faucets. As examples of such measurements, in figure
II.6 we show the fluctuation function calculated from the same simulated
time series as presented in figure II.5. It is clear that for all the
fluxes the Hurst exponent is equal to zero. Actually, the authors found
$\alpha = 0.00$ [4] for all tested cases, within much larger time series,
both experimental and simulated, with at least 10-fold larger accuracy
than for the case of heart beats [33,34].

	Another property which is the same for the time series of drops
and heart beats is the probability distribution of the intervals $B(n)$
themselves. For both systems, it was shown [33-4] that it is well
described by a L\'evy distribution

$$P(I) = (1/\pi) \int_0^\infty \exp(\gamma q^\psi) \cos(qI)
dq \,\,\,\, , \eqno(4)$$

\noindent where $I = B\, - <B>_n$. For both systems one gets $\psi \cong
1.7$ ($\psi = 2$ would correspond to a Gaussian). The exponent $\psi < 2$
gives rise to a fat tail on the distribution plot, according to a
power-law instead of the normal Gaussian exponential tail. The unimportant
coefficient $\gamma$ simply defines the distribution width. Note that
$\alpha$ and $\psi$ are independent exponents: $\alpha$ describes the
particular chronology of events (it changes to $\alpha = 1/2$ under data
shuffling), whereas $\psi$ describes their intensities (does not change
under shuffling). All these properties were obtained for different
experimental and simulational conditions. Further comparisons of the leaky
faucet and heart beat series were made by using other quantities, e.g.
entropies and higher dimensional return maps [5]. The results suggested
that a common physical mechanism, leading to long-term anticorrelations,
could be present in both systems.

	As mentioned in the beginning of this section, nobody knows if the
leaky faucet is really chaotic. Up to now, different calculations have
shown that the Lyapunov exponent could also be zero, but the precision is
still unsatisfactory. This region is particularly difficult to study since
the majority of methods available for Lyapunov exponent calculation assume
exponential separation of orbits. However, the separation is not
exponential if $\lambda = 0$. The suggestion that the Lyapunov exponent
could be zero is mainly based on the fact that both leaky faucets and
healthy hearts present long-term memory features, while any non-vanishing
Lyapunov exponent $\lambda \not= 0$ would set a characteristic {\bf
finite} transient time $1/|\lambda|$ for the underlying dynamics.
Moreover, as usual for biological systems, a healthy heart could follow a
critical dynamic in the ``edge of chaos'' [33,34,11], i.e. with Lyapunov
exponent $\lambda = 0$ for which the transient time diverges to {\bf
infinity}. This feature gives advantages in the adaptation of the system
to unknown environmental changes [33-35]. The lattice drop model described
here can be very useful to understand such properties. It allows one to
generate very large time series in a fast and controllable way.

\vskip 1cm
{\bf III -- Nuclear Multifragmentation}\parG

	The last two decades have seen an increasing interest in the
experimental and theoretical study of heavy ions in collision. A major
reason for this outburst is the expectation that at high enough energies,
of the order of $200$ MeV per nucleon, this system could probe the region
in phase space where a transition liquid-gas has been predicted by a
number of models for infinite nuclear matter [36]. In fact, the observed
distribution of mass (or charge) of the intermediate mass fragments (IMF)
generated by this process shows scale invariance along several decades; it
has been claimed that the ubiquitous presence of this power-law for a
number of systems of different compositions is a remnant, as observed in a
finite system, of the above-mentioned transition. It has become standard
practice in the literature to call nuclear multifragmentation the physical
process that underlies these fractal and correlated distributions of IMF.  
Multifragmentation is then understood as the nuclear matter analog for the
critical opalescence phenomenon observed in normal fluids.

	As a theoretical support for this claim, the archetypical
continuous transition in a percolation model has been examined. It could
be shown, from simulational work, that the scale invariance for the
distribution of cluster sizes present in the thermodynamical limit has a
counterpart in finite, and small, systems [37]. Moreover, the correlations
between the moments of various orders of this distribution, also present
in this limit, still survive in the latter systems. Analysis of nuclear
collision experiments, made on an event-by-event basis, have shown a
striking parallel with these numerical results on the percolation of
small-sized systems. In particular, the critical exponents associated with
the percolating transition, as revealed in small systems, appear to be,
within errors, also valid for nuclear multifragmentation. Other than the
exponent $\tau$ associated to the scale invariant distribution of IMF,
defined through the equation

$$n(s,T) \sim s^{- \tau} f(T)\,\,\,\, ,\eqno(5)$$

\noindent it has been claimed that the exponent $\gamma$ according to
which the susceptibility diverges in the infinite systems, defined through

$$M_2(T) \sim (T - T_c)^{- \gamma}\,\,\,\, ,\eqno(6)$$ 

\noindent could also be determined from the experimental data, and that
its value is again consistent with the percolation transition [38]. In the
above equations, $n(s,T)$ is the distribution of IMF as a function of size
$s$, $M_2(T)$ is the second moment of this distribution

$$M_k(T) = \sum_s s^k n(s,T)\,\,\,\, ,\eqno(7)$$

\noindent $T$ is the temperature of the system and $T_c$ its critical
value.

	These results seem to support the analogy between percolation and
nuclear multifragmentation, but are somewhat puzzling. Percolation is a
geometrical and equilibrium continuous transition; on the other hand, it
is not at all clear, and perhaps even not true at all, that the
multifragmenting nuclei can be treated as an equilibrium system. One must
add that a continuous transition needs fine tuning of at least one
relevant scaling field, and multifragmentation appears to be present in a
large range of different beam energies. In fact, the liquid-gas transition
to which multifragmentation is referred is continuous only in a subspace
of zero measure of phase space. These comments should suffice to justify
an alternative approach to the problem that takes into account its
intrinsic dynamical nature and that makes no {\it a priori} assumption on
the order of the transition.

	Dynamical models have been used recently to address
multifragmentation, with some success; we mention here those based on
quantum molecular dynamics [39], on cellular automata inspired by these
dynamics [40], and on the lattice gas model [41]. Our approach to the
problem bears some formal resemblance to the latter two; we model the
compound intermediate nucleus at the freeze-out density as a lattice drop,
a compact 3-D cluster of Ising spins pointing $up$, representing the
nucleons, surrounded by spins pointing $down$, as mentioned in the
introduction. The basic assumptions of the model can be summarized as:

$\bullet$ The entrance and exit channels are decoupled;

$\bullet$ The fragments come from the decay of a thermodynamically 
equilibrated, but still excited, source;

$\bullet$ Decay proceeds through surface rearrangements of the nucleons that 
compose the source, and through bulk radiative processes. Multifragmentation 
is the outcome of surface relaxation.

	The dynamical evolution of the system starts at some chosen
initial temperature, related to the excitation energy of the system. The
subsequent motion in phase space follows the general rules stated in the
introduction, with an Ising Hamiltonian that includes coupling with
nearest and next-nearest neighbours (hereafter, $g = 0$). In this case,
the flipping of spins is accepted through a Metropolis algorithm. The
result of these update rules is a Kawasaki-like dynamics, with
conservation of ``magnetisation,'' or mass (number of nucleons), which
causes a continuous reshaping of the surface of the system. One can
associate this dynamics to a canonical walk on the landscape generated by
the surface part of the system's free energy. After each successful
flipping, the (surface) energy variation that it entails is recorded and
the resulting drop geometry is examined for compactness. If it has become
two disjoint blobs, the smallest one is called a fragment, its mass
included in a statistical distribution, and it is erased from the rest of
the simulation; the accumulated variation in the system's energy since the
last fragment was formed is then taken into account, resulting in a new
value for the excitation energy of the nucleus. Other than this surface
relaxation process, a bulk radiative decay is also simulated. After each
fixed number of Monte Carlo time steps, the excitation energy is decreased
by a constant factor, generating an exponential decay.  The combination of
these two parameters represent in fact a ratio between the intensity of
surface and bulk processes, and play an interesting role in what follows.
A complete description of the model can be found in [8], together with
comments on its computer implementation.

	An important final detail relates to the way in which the running
value of the excitation energy generates a thermodynamic temperature for
the system. This relation is usually called the caloric curve in the
nuclear physics community, and its actual shape or analytical expression
has been the subject of intense study and controversy in recent years. In
our simulations, we used two different forms for this caloric relation, a
Boltzmann gas-like linear relation $T \sim E$ and a quadratic Fermi
gas-like $T \sim \sqrt E$.

	We present in figure III.1 the correlations between moments of
higher orders of the fragment distribution seen both in the experiments
and in our simulations. For a distribution satisfying a static scaling
{\it ansatz} these correlations can be derived analytically and read in
general

$$\log(M_k(T)) \sim \lambda_{k/j}\, \log(M_j(T))\,\,\,\, .\eqno(8)$$

	The slope $\lambda_{k/j}$ of these $log \times log$ plots can be
related to the Fisher exponent $\tau$, and have to satisfy a consistency
equation near criticality, as shown in [37]:

$$\tau = (j + 1) - (k - j)/(\lambda_{k/j} - 1), \ \ 
(k' - j) \lambda_{k/j} - (k - j) \lambda_{k'/j} = k' - k\,\,\,\, .
\eqno(9)$$

\noindent The consistency equation can be used in the simulations to
identify the realisations that most closely follow the correlations
expected in the thermodynamical limit, and from these one can infer the
value for $\tau$. Proceeding in this manner, we found $\tau = 2.18$, the
same for both caloric curves and in accordance with the value inferred
from the experiments.

	For the determination of a second critical exponent we followed
guidelines put forth in [38] when dealing with the same problem for the
experimental data. The central idea is to explore the equality of the
exponent $\gamma$ for the second moment of the distribution when
calculated above and below the transition. As an interesting by-product,
this technique enables a more precise determination of the transition
point. One proceeds in analogy with percolation in finite lattices, where
the order parameter is the density of the percolating, or largest,
cluster. This cluster is not considered for the second moment of the
distribution in the liquid phase, or below the transition, where it is the
seed for the nucleation of the stable phase. For the gas phase, above the
transition, all clusters are considered. By a trial and error procedure,
these moments are calculated for each candidate to the transition point in
a certain range, and the slopes of a $log \times log$ plot as a function
of the parameter measuring the distance to the transition point are
compared. A best match between the two is the criterion that determines
the critical point, and $\gamma$ is also immediately available. Figure
III.2 illustrates the procedure and shows the best determination for both
the transition point and the exponent $\gamma$ for a Fermi-gas caloric
curve and a particular value for the surface-to-bulk time scale ratio
$a=0.998$. The value found, $\gamma = 1.80$, is again consistent with the
conjecture that puts nuclear multifragmentation in the universality class
of $3D$ percolation. Contrary to the robust value of $\tau =2.18$, it was
also found that these last choices ($T \sim \sqrt E$ and $a=0.998$) have a
definite impact on the exponent $\gamma$. Although also found in solvable
models such as the eight vertices model [42], where it can be related to a
marginal scaling field, this rather unusual continuous dependence of a
critical exponent on a parameter is worth mentioning. In particular, a
conjecture [43], relating the controversy among experimental groups about
the value of $\gamma$ to the particular pair target-beam used and the
resulting differences in surface-to-bulk relaxation time scales, appeared
in the literature shortly after our result was first published [9].

	To summarize, we have shown that simulations of a lattice model,
evolving through dissipative surface and bulk relaxation, with Metropolis
dynamical rules and a short-range Ising Hamiltonian, can reproduce both
qualitative and quantitative aspects observed in nuclear
multifragmentation. This agreement supports the claim that the relevant
physics of the processes involved does not depend on the microscopics of
the interactions, which would be in any case unlikely to be derived from
first principles, and to which one has only indirect access, if any. It is
not incidental that a pathway like ours can suggest and support new
relevant physics, such as pointed out in the last paragraph: by focusing
on a few well chosen features of the phenomena being studied, the role of
each can be more adequately explored.  This newly acquired knowledge will
in turn support the choices to be made in the next generation of models,
in a fast growing evolutionary chain made possible by the simulational
approach.

\vskip 1cm
{\bf IV -- Mercury Drops}\parG

	A nice experiment was introduced in [44]. A mercury drop falls
from a certain height on the floor. As a consequence of the collision,
fragments are formed. By classifying these fragments by size (or mass)
according to a logarithmic scale (bins), and counting how many of them
exist within each class, one can measure the size distribution
probability. Good statistics would be achieved after many repetitions of
the experiment, starting with the same initial drop as well as the same
falling height, superimposing all countings at the end. This process is
similar to the nuclear multifragmentation experiment described in the
previous section. One can also predict a similar behaviour: a broad
distribution of fragment sizes could appear, perhaps following some
power-law.

	The outcome of the experiment [44] is not a simple power-law. By
plotting the countings according to a $log \times log$ scale, one has
indeed a broad distribution of fragment sizes, but not a single straight
line. Instead, the experimental data is better fitted by two straight
lines (see figure IV.1, to be described later). A crossover point in
between the small- and large-fragment regimes appears as a rounded knee on
this experimental plot. The drop-counting strategy adopted by the authors
of [44] is a hard one: each fragment is measured through a microscope.
That is why their experimental data do not present a very good statistics,
and a precise definition of the mathematical form of the distribution
cannot be inferred, moreover for the small-drop limit. Contrary to figure
IV.1, the left part of the corresponding plot presented in [44] seems to
approach a horizontal line, indicating some sort of counting saturation
for smaller and smaller fragments. We credit this feature to an
underestimated counting of small fragments at the microscope, not to a
real saturation effect.

	That is why we decided to repeat the experiment following an
alternative strategy: we take a photograph of the fragments, which is then
digitalised by using a scanner. The fragment counting was then performed
by a computer program. This procedure gives us confidence in the
small-drop counting, as compared with [44]. Our first, preliminary results
show the same behaviour as in [44] for large fragments, and are displayed
in figure IV.1. It corresponds to two different falling heights $h =
220\,$cm and $240\,$cm, with 10 realisations of the experiment for each
height. Once we used real photographic paper, with a much higher
resolution than the scanner, our experiment is limited only by the latter.
Thus, it is possible to improve our results by including smaller yet drops
into the statistics (figure IV.1 was prepared with a low scanner
resolution, due to limitations of memory and time on our outdated
computer). For the present work, nevertheless, the preliminary results of
figure IV.1, already with a much better counting of small fragments as
compared with [44], are enough in order to test our dynamic lattice drop
model as applied to this problem. Experimental improvements are in course
and will be presented elsewhere.

	Let's describe now our dynamic model. The computer simulations
starts at the very moment when the initial mercury drop reaches the floor.
It is represented by a round cluster of $N$ black pixels (spins $up$) on a
square or cubic lattice, just touching the floor represented by an inert
bottom boundary. The total kinetic energy $E$, proportional to the falling
height already traced downwards, is given since the beginning. Being a
coherent kinetic energy before the crash, it is shared among all drop
pixels, each one carrying an amount of $e = E/N$. The temperature is set
to $T = 0$, once one has no random motion yet. From this initial
situation, the collision with the floor will be performed as follows, step
by step: $i$) the $n$ pixels currently touching the floor are transferred
to random positions at the current free surface, and the drop as a whole
is moved one row down; $ii$) the energy $e$ carried by each one of these
$n$ pixels is set to 0, decreasing the coherent kinetic energy of the
whole drop; $iii$) the same decrement $\Delta{E}$ is transformed in
incoherent, thermal energy, by increasing the temperature from $T$ to $T +
\Delta{E}$; $iv$) The drop shape is then allowed to relax, by performing
$r$ movements exchanging the positions of a spin $up$ with another spin
$down$, both randomly chosen at the current free surface, according to the
Metropolis rule under temperature $T$ for the first-and-second neighbour
Ising Hamiltonian. This last step is the same already used before for the
leaky faucet and nuclear multifragmentation. The four steps are repeated
iteratively.

	At some moment during this process, some piece of the drop becomes
disconnected from the main part still touching the floor. This piece is
considered a fragment. If, due to this fragmentation, the energy of the
whole system increases, then this fragment is discarded and counted into
the statistics. Also in this case, the temperature $T$ is decreased to $T
- k E_f$, where $E_f$ is the surface energy carried out by the fragment.
The dynamics described in the last paragraph goes on, further relaxing the
remainder drop. The process stops when the drop vanishes, or when the
temperature is low enough to avoid any further activity. The whole process
is performed again and again, starting always from the same initial drop
with the same initial energy, in order to accumulate a good statistics. At
the end, the fragments are classified by their sizes (mass), according to
a logarithmic scale (bins), taking into account all realisations performed
with the same initial energy.

	Figures IV.2 and 3 show two typical distributions obtained by our
model, to be compared with the experimental couterparts in figure IV.1.
The general behaviour with two power-law regimes, one for small and the
other for large fragments, is obtained always for a lot of different
choices for the set of parameters $E$, $r$ and $k$, both for the square
and the cubic lattice. Note that our simulational statistics is much
higher than the experimental one. In spite of the striking similarity
between these simulational results and the experimental observations, at
the present moment we consider our model still very crude in order to
allow any quantitative comparison. Both our experiments and simulations
are still in a preliminary stage. Anyway, looking to Figures IV.1, 2 and
3, we can hope our naive model could be useful in understanding the
addressed physical problem. At least two points are still missing: $i$) a
better definition of the experimental curve, figure IV.1, by improving
both the resolution as well as the statistics; $ii$) a theoretical
explanation for the crossover between the small and large fragment
regimes, which is already evident in both our experiment and simulation.

\vskip 1cm
{\bf V -- Magnetic Hysteresis}\parG

	A nice example of non-equilibrium physical behaviour is the
hysteresis curve, magnetisation $versus$ applied magnetic field. Consider
a virgin sample of some ferromagnetic material below its Curie critical
temperature. Virgin means that the sample is not magnetised yet, i.e. it
was never exposed to any external magnetic field. One can obtain such a
sample simply by heating it far above the Curie temperature, and then
allowing it to return back to a lower temperature, free from any external
magnetic field. In this case, inside each magnetic grain of the sample one
can find different magnetic domains, separated from each other by domain
walls. As the magnetic axis inside each domain points in a random
direction, the whole grain (and thus also the whole sample) presents a
macroscopically vanishing magnetisation.

	Consider now the sample kept under a fixed temperature, below the
Curie point. Applying a small external magnetic field, some few domains
already parallel to the field are enlarged, at the expenses of other
neighbouring domains which shrink. As a result, the sample presents now a
small macroscopic magnetisation, far below its maximum possible saturation
value. This occurs because other domains, for which the external field is
not strong enough to provide the energy necessary to jump the barrier
corresponding to flipping the spins near the domain wall, remain as they
were before the external field was applied. By increasing a little bit the
external field, the whole magnetisation is supposed also to increase
proportionally to the (still small) field. Thus, the plot magnetisation
$versus$ field follows more or less a straight line, starting from the
origin (see the beginning of the curve in figure V.7, just below the
symbol 0, to be explained later).

	However, this linear dependence between magnetisation and external
field cannot continue for larger values of the field. Experimentally, as
the external field is gradually increased, one observes first a change in
the slope, the magnetisation increasing faster and faster in response to
the applied field. Later, for high enough values of the external field,
the magnetisation eventually saturates, all magnetic domains of the sample
pointing in the same imposed direction (see branch 0 of figure V.7).

	The explanation for this behaviour is simple. Some domain walls,
particularly where impurities are absent, can easily be deformed just by
transferring a spin from one magnetic domain to another neighbouring one
(i.e. a domain wall movement). A simple example is shown in figure V.1,
where a $64 \times 32$ ``grain'' of Ising spins displays two magnetic
domains. No external field is yet applied, and the magnetisation still
vanishes. Spins near the domain wall are easier to be flipped than other
spins deep inside each domain. Along the wall, however, spins near the
grain boundary, i.e. those touching the upper or the lower row in figure
V.1, are not so easy to be flipped, because they are pinned by the
non-magnetic material outside the grain. Thus, by applying a small
magnetic external field, the result is a round domain wall, as shown in
figures V.2 and V.3, instead of the flat one in figure V.1. The
magnetisation increases, in response to the applied field, but not so
much.

	Within our simple model, in order to mimic this pinning effect, we
adopted the following boundary conditions for spins at the upper or lower
rows. For these sites, instead of 8 neigbouring spins (first and second
neighbours on the square lattice), one has only 5. The 3 missing
neighbours are then supposed to point in the same direction of the spin
currently being updated. Now, instead of choosing a pair of spins, one
$up$ and the other $down$, one chooses only a random one, $up$ or $down$,
along the domain wall. It will be flipped or not, according to the
canonical Boltzmann weight $\exp(-\Delta{E}/T)$, where $\Delta{E}$ is the
energy increment due to flipping. In order to compute this energy jump
$\Delta{E}$, besides the external field, one takes into account the Ising
coupling energy between the tossed spin and its 8 fixed neighbours. Would
this spin be tossed at the grain boundaries, the 3 missing neigbours
(supposed to be parallel to it) would give always a positive contribution
to $\Delta{E}$, thus difficulting the flipping as compared to other spins
along the domain wall. Figures V.1 $\dots$ 6 show snapshots of the Ising
spins on our $64 \times 32$ grain, for increasing values of the external
magnetic field $h$. The magnetisation $m$ increases slowly for figures
V.1, 2 and 3, for which the domain wall remains pinned at the grain
boundaries: it is bent but does not slide yet. In figure V.4, however, the
external field is already strong enough to force the domain wall to slide
against the boundary pinning: the magnetisation starts to increase
according to a larger rate, as in figure V.5. Finally, figure V.6 shows
the situation where the magnetisation is already saturated. The complete
plot magnetisation $versus$ field corresponds to the branch denoted by 0
in figure V.7. The field is gradually increased, in steps of $\Delta{h} =
0.01$. In between two successive field upgrades, $M = 100$ spin flips are
tried along the current domain wall.

	The branch denoted by 1 in figure V.7 corresponds to decreasing
the field back, following the same steps. Only for already negative values
of the field, the magnetisation starts to decrease again from its
saturation value. Eventually, after crossing the value 0 at the so-called
coercitive field, it becomes again parallel to the external field, and
saturates for large enough negative values of the field. The last branch
denoted by 2 in figure V.7 corresponds to another field sweep, again step
by step, now from negative to positive values. One can obtain different
hysteresis curves like that, by changing the values of $M$ and/or
$\Delta{h}$, as well as the fixed temperature $T$ adopted in order to
decide to flip or not the tossed spin.

\vskip 1cm
{\bf VI -- Annealing Effects in Multilayers}\parG

	Using the ideas introduced in the previous sections, the same
model was adopted [12] to study the annealing effects on multilayers. Such
study was motived by the experimental fact that electronic properties of
multilayered systems can be strongly affected by the quality of their
interfaces. A couple of examples are the giant magnetoresistance effect
(GMR) and the interlayer magnetic coupling [45-47]. Therefore, the
control of interfacial roughness and interdiffusion is an important issue
both for basic research and for the use of those systems in electronic
devices.

	In that study [12], a two-dimensional system made of immiscible
materials A and B was considered, consisting of a stripe of material A
sandwiched by B. $L$ and $H$ represent the lateral dimension and height of
the whole system, respectively, and $W$ is the mean width of stripe $A$
(see figure VI.1). An example of real system composed by two immiscible
materials arranged in such a way are Co/Ag multilayers (see [48] and
references therein). For this system it was suggested that the interfacial
roughness displays a minimum as a function of the annealing temperature.
It has also been found that relatively thin Co layers may break, forming
clusters, for sufficiently high annealing temperatures [48].

	Each spin $S_i$ of our model takes the values $\pm 1$, depending
on whether the site $i$ is occupied by an A ($+1$) or B ($-1$) atom. The
Ising energy summation runs {\bf over pairs of nearest neighbours only}
and the interaction favours agglomeration of atoms of the same type. Now,
the role of temperature is crucial. The system is placed in contact with a
heat bath at a temperature $T$, and a spin-flip (AB interchange) Monte
Carlo dynamics is performed by Kawasaki trials, where pair updates are
made, conserving both the numbers of A and B atoms [49]. This is executed
by randomly choosing one site A at the interface, and interchanging it
with a {\bf nearest neighbour B atom, also picked at random}. Such
interchange preserves the total number of atoms of each species, but may
change the total energy of the system by $\Delta{E}$. The Metropolis
algorithm is then used to define the acceptable interchange trials,
according to the ratio $\Delta E/T$, where the Boltzmann's constant is set
to unity. The first and last rows are fixed with $W$ adjacent A atoms
each, face to face, in order to avoid lateral motion of the stripe of A
atoms. This could be interpreted as a surface effect.

	The annealing process is simulated as follows: we take a system
with randomized interfaces, in which layer A has mean width $W$ with
maximum mean square deviation $\delta$. The system is then thermalized at
a given temperature $T_a$, allowing for $n_a$ interface relaxations. The
temperature is gradually reduced, in steps $dT$, until it reaches a value
$T_0$. After each temperature reduction, $n_T$ interface relaxations are
performed. The parameter $n_a$ regulates the amount of time the sample was
kept at temperature $T_a$, and $n_T$ controls the rate at which the
temperature is reduced from $T_a$ to $T_0$. Roughness is measured by
computing $\sigma = \sum_{<ij>} (1-S_{i}S_{j})/2 - 2H/a$, where $a$ is the
lattice constant. Clearly, for perfectly flat interfaces one has $\sigma =
0$, and it increases with A-B interdiffusion.

	For this problem, the interest is to calculate the mean roughness
variation $\langle(\sigma_f - \sigma_i)/\sigma_i \rangle$, as a function
of the annealing temperature $T_a$, for several widths $W$. Here $\langle
...  \rangle$ represents an ensemble average over $N$ samples, whereas
$\sigma_i$ and $\sigma_f$ are the corresponding values of $\sigma$ before
and after the annealing process is executed. Some results are shown in
figure VI.2. Lengths are measured in units of $a$. The calculations were
made for $N = 100$ different samples, all with $H = 150$, $L = 40+W$, and
$\delta = 2$. We assume that $T_a$ is reduced down to $T_0 = 0.01$ in
steps of $dT = 1/32$, with $n_a = 20$ and $n_T = 10$. It is evident from
figure VI.2 that the mean roughness variation has a minimum, as
experimentally inferred. Its position does not depend much on $W$, and we
have also verified that for different values of $H$, $L$, $n_a$, $n_T$ and
$\delta$ (provided that $\delta \ll W$), the results follow the same
behaviour, all showing a clear minimum at $T_a \cong 1.2$. The insets
depict snapshots of the final interface configuration at $T_0$ for some
values of $T_a$. Is is clear that for certain values of $W$ and high
enough values of $T_a$, the A layer may break into clusters. This effect
can also be studied by this kind of model.

	It was also shown [12] that some quantities as the average number
of A atoms having $z$ first and second nearest neighbours of type B, for
different values of $T_a$, provide results similar to the experimental
ones obtained by nuclear magnetic resonance experiments [50]. Since this
simple model is able to produce layers with controllable roughness
(determined by the annealing temperature) it can be very useful to study
transport properties on multilayers. Work along these lines is in
progress.

\vskip 1cm
{\bf VII -- Conclusions}\parG

	Critical equilibrium situations occur in many systems in Nature.
Generally, by varying some external control parameter $T$ (temperature,
for instance) the system can suffer a continuous phase transition when
some critical value $T_c$ is surpassed. By keeping the system in
equilibrium just at $T_c$, its behaviour becomes completely distinct from
that observed both above and below $T_c$. Instead of the normal
exponential decaying of spatial correlations (in general) observed by
increasing the distance between two points randomly picked inside the
systems, just at $T_c$ these same correlations decay as power-laws. The
exponential decaying feature provides a well-defined length scale $\xi(T)$
for the system at $T\not=T_c$: two different positions separated by a
distance larger than $\xi(T)$ barely influence each other. One can
consider the system as composed by finite pieces with typical length
$\xi(T)$ each: the behaviour of the system as a whole is simply obtained
by adding the extensive quantities (energy, mass, magnetisation, etc)
corresponding to its component pieces treated separated from each other.
The system is linear, in the sense that the whole is the sum of the parts.

	On the other hand, just at $T_c$, the lack of an exponential
decaying gives rise to an equivalent lack of any characteristic length
scale: the system no longer behaves as the linear superposition of finite
pieces, and all length scales must be taken into account. The adequate
theory to treat such singular, critical equilibrium situations was awarded
with the 1982 Nobel prize for Kenneth Wilson (see [51] for a friendly
introduction). According to it, instead of the linear behaviour,
successively performed length scaling operations, i.e. successive zooms
multiplying by $\lambda$ the unit length of its component pieces, lead to
successive multiplications of some quantity $Q$ by $\lambda^{\phi_Q}$,
where the exponent $\phi_Q$ is characteristic of each quantity $Q$. Note
that a power-law relation between $Q$ and the length $L$ is invariant
under such an endless scaling procedure, but not an exponential decaying
relation which would fast vanish, instead.

	The distinguishing feature of critical situations is the
universality: the same set of critical exponents $\phi_Q$ for the various
quantities of interest holds for entire classes of different systems and
models. Only some general characteristics of the system, namely its
geometrical dimension and the symmetries of its order parameter suffice to
define its critical equilibrium behaviour. For instance, in spite of the
quantum character and the complicated interactions between neighbouring
molecules, the critical behaviour of water at $T_c = 374^{\rm o}$C can be
modeled by a simple three-dimensional Ising ferromagnet! This is only a
striking example among a huge set of others where universality is
observed. That is why models very simplified at the microscopic level can
be applied with success to study the critical equilibrium behaviour of a
much more complicated real system. These lines were extensively followed
during the last four decades, within the fast growing research field of
critical equilibrium phenomena.

	Concerning non-equilibrium systems, too, examples of critical
dynamics are numerous. In this case, a power-law decay behaviour in {\bf
time}, besides length, also appears. This means that the dynamic evolution
of such systems is insensitive to {\bf short-term} simplifications,
besides the microscopic details above mentioned. A striking example of
this robustness was shown in [27], where the leaky faucet experiment was
performed with clusters of ants pending from a thin vertical rope, insted
of water drops. The ants go by themselves to this rope, one after the
other, forming a growing cluster. Suddently, a drop of them falls, but
some few ants still remain pending from the rope, giving rise to a new
growing drop. The ``short-range coupling'' between ants glued on each
other has nothing to do with the van der Waals attraction between water
molecules, neither with the simplified coupling between neighbouring Ising
spins. Nevertheless, exactly the same results described in section II were
also obtained in [27], with ants. In the present work, we have exhibited
some examples where a simple drop dynamic model reproduces very well the
behaviour of real systems. The common feature among these otherwise
completely distinct systems is the fact that they all present {\bf
long-term} memory features, allowing us to apply our very simplified
model.

\vskip 1cm
{\bf Acknowledgements}\parG

	We are indebted to J\"urgen Stilck and Dietrich Stauffer for
critical readings of the manuscript. This work was partially supported by
Brazilian agencies CAPES, CNPq and FAPERJ. J.S.S.M. is supported as a
Visiting Fellow by CIRES, University of Colorado at Boulder.

\vskip 1cm
{\bf References}\parG

\item{[1]} S.S. Manna, H.J. Herrmann and D.P. Landau, {\it J. Stat. Phys.}
{\bf 66}, 1155 (1992).\par

\item{[2]} P.M.C. de Oliveira and T.J.P. Penna, {\it J. Stat. Phys.} {\bf
73}, 789 (1993).\par

\item{[3]} P.M.C. de Oliveira and T.J.P. Penna, {\it Int. J. Mod. Phys.}
{\bf C5}, 997 (1994).\par

\item{[4]} T.J.P. Penna, P.M.C. de Oliveira, J.C. Sartorelli, W.M. Gon\c
calves and R.D. Pinto, {\it Phys. Rev.} {\bf E52}, R2168 (1995).\par

\item{[5]} A.R. Lima, T.J.P. Penna and P.M.C. de Oliveira, {\it Int. J.
Mod. Phys.} {\bf C8}, 1073 (1997); A.R. Lima, {\it MSc. Thesis},
Universidade Federal Fluminense, Niter\'oi, Brazil (1997).\par

\item{[6]} P.M.C. de Oliveira, J.S.S. Martins and A.S. de Toledo, {\it
Phys. Rev.} {\bf C55}, 3174 (1997).\par

\item{[7]} P.M.C. de Oliveira and A.P. Guimar\~aes, {\sl Monte Carlo
Simulations of Hysteresis Loops}, International Conference on Magnetism
(1997), Cairns, Australia (unpublished).\par

\item{[8]} J.S.S. Martins and P.M.C. de Oliveira, {\it Int. J. Mod. Phys.}
{\bf C9}, 867 (1998).\par

\item{[9]} J.S.S. Martins and P.M.C. de Oliveira, {\it Nucl. Phys.} {\bf
A643}, 433 (1998).\par

\item{[10]} P.M.C. de Oliveira, J.S.S. Martins, C. Moukarzel and
C.A.F. Leite, {\sl Falling Mercury Drops}, Workshop on Complex Systems
(1998), Bras\'\i lia, Brazil (unpublished).\par

\item{[11]} S. Moss de Oliveira, P.M.C. de Oliveira and D. Stauffer, {\sl
Evolution, Money, War and Computers: Non-Traditional Applications of
Computational Statistical Physics}, section 3.3, B.G. Teubner, Stuttgart
Leipzig, ISBN 3-519-00279-5 (1999).\par

\item{[12]} A.R. Lima, M.S. Ferreira, J. d'Albuquerque e Castro and R.B.
Muniz, to appear in {\it J. Magn. Magn. Mater.} (2000).

\item{[13]} R. Shaw, {\sl The Dripping Faucet as a Model Chaotic System},
Aerial Press, Santa Cruz, California (1984).\par

\item{[14]} P. Martien, S.C. Pope, P.L. Scott and R. Shaw, {\it Phys.
Lett.} {\bf 110A}, 339 (1985).\par

\item{[15]} H.N.N. Y\'epes, A.L.S. Brito, C.A. Vargas and L.A. Vicente,
{\it Eur. J. Phys.} {\bf 10}, 99 (1989).\par

\item{[16]} X. Wu and A. Schelly, {\it Physica} {\bf D40}, 433 (1989).\par

\item{[17]} K. Dreyer and F.R. Hickey, {\it Am. J. Phys.} {\bf 59}, 619
(1991).\par

\item{[18]} J. Austin, {\it Phys. Lett.} {\bf 115A}, 148 (1991).\par

\item{[19]} X.D. Shi, M.P. Brenner, and S.R. Nagel, {\it Science} {\bf
265}, 219 (1994).\par

\item{[20]} J.C. Sartorelli, W.M. Gon\c calves and R.D. Pinto, {\it Phys.
Rev.} {\bf E49}, 3963 (1994).\par

\item{[21]} R.D. Pinto, W.M. Gon\c calves, J.C. Sartorelli and M.J. de
Oliveira, {\it Phys. Rev.} {\bf E52}, 6896 (1995).\par

\item{[22]} G.I.S. Ortiz and A.L.S. Brito {\it Phys. Lett.} {\bf 203A},
300 (1995).\par

\item{[23]} M.S.F. da Rocha, J.C. Sartorelli, W.M. Gon\c calves and R.D.
Pinto, {\it Phys. Rev.} {\bf E54}, 2378 (1996); M.S.F. da Rocha, {\it MSc.
Thesis}, Universidade de S\~ao Paulo, S\~ao Paulo, Brazil (1996).\par

\item{[24]} J.G. Marques da Silva, J.C. Sartorelli, W.M. Gon\c calves and
R.D. Pinto, {\it Phys. Lett.} {\bf 226A}, 269 (1997).\par

\item{[25]} W.M. Gon\c calves, R.D. Pinto, J.C. Sartorelli and M.J. de
Oliveira, {\it Physica} {\bf A257}, 385 (1998).\par

\item{[26]} R.D. Pinto, W.M. Gon\c calves, J.C. Sartorelli, I.L. Caldas
and M.S. Baptista, {\it Phys. Rev.} {\bf E58}, 4009 (1998).\par

\item{[27]} E. Bonabeau, G. Theraulaz, J.-L. Deneubourg, A. Lioni, F.
Libert, C. Sauwens and L. Passera, {\it Phys. Rev.} {\bf E57}, 5904
(1998).\par

\item{[28]} S.R. Nagel, {\it Am. J. Phys.} {\bf 67}, 17 (1999).\par

\item{[29]} G.I. S\'anchez-Ortiz and A. L. Salas-Brito, {\it Physica} {\bf
D89}, 151 (1995).\par

\item{[30]} A. d'Innocenzo and L. Renna, {\it Phys. Lett.} {\bf 220A}, 75
(1996).\par

\item{[31]} N. Fuchikami, S. Ishioka and K. Kiyono, {\it J. Phys. Soc.
Jpn.} {\bf 68}, 1185 (1999).\par

\item{[32]} K. Kiyono and N. Fuchikami, {\it J. Phys. Soc. Jpn.} {\bf 68},
3259 (1999).\par

\item{[33]} C.-K. Peng, J. Mietus, J.M. Haussdorf, S. Havlin, H.E. Stanley 
and A.L. Goldberger, {\it Phys. Rev. Lett.} {\bf 70}, 1343 (1993).\par

\item{[34]} H.E. Stanley, S.V. Buldyrev, A.L. Goldeberger, Z.D.
Goldberger, S. Havlin, R.N. Mategna, S.M. Ossadnik, C.-K. Peng and M.
Simons, {\it Physica} {\bf A205}, 214 (1994).\par

\item{[35]} P.M.C. de Oliveira, to appear in {\it Theor. Biosci.} {\bf
120} (2001).\par

\item{[36]} A.L. Goodman, J.I. Kapusta and A.Z. Mekjian, {\it Phys. Rev.} 
{\bf C30}, 851 (1984).\par

\item{[37]} X. Campi, {\it J. Phys.} {\bf A19}, L917 (1986); for
percolation itself, see D. Stauffer and A. Aharony {\sl Introduction to
Percolation Theory}, Taylor and Francis, London (1994).\par

\item{[38]} M.L. Gilkes et al, {\it Phys. Rev. Lett.} {\bf 73}, 1590
(1994); J.B. Elliot et al, {\it Phys. Rev.} {\bf C55}, 1319 (1997).\par

\item{[39]} J. Aichelin, {\it Phys. Rep.} {\bf 202}, 233 (1991).\par

\item{[40]} A. Lejeune, J. Perdang and J. Richert, {\it Phys. Rev.} {\bf
E60}, 2601 (1999).\par

\item{[41]} X. Campi and H. Krivine, {\it Nucl. Phys.} {\bf A620}, 46
(1997); S.K. Samaddar and S. Das Gupta, {\it Phys. Rev.} {\bf C61}, 034610
(2000).\par

\item{[42]} R.J. Baxter, {\sl Exactly Solved Models in Statistical
Mechanics}, Academic Press (1989).\par

\item{[43]} L. Beaulieu et al, {\it Phys. Rev. Lett.} {\bf 84}, 5971
(2000); S. Das Gupta, A. Majumder, S. Pratt and A. Mekjian, pre-print 
nucl-th/9903007.\par

\item{[44]} O. Sotolongo-Costa, Y. Moreno-Vega, J.J. Lloveras-Gonz\'alez
and J.C. Antoranz, {\it Phys. Rev. Lett.} {\bf 76}, 42 (1996).\par

\item{[45]} J. Unguris, R.J. Celotta,  and D.T. Pierce, {\it Phys. Rev.
Lett.} {bf 67}, 140 (1991).\par

\item{[46]} B. Heinrich and J.A.C. Bland, {\sl Ultrathin Magnetic
Structures} (Springer, Berlin, 1994), Vol. 2.\par

\item{[47]} B. Dieny, {\it J. Magn. Magn. Mater.} {\bf 136}, 335
(1994).\par

\item{[48]} L.F. Schelp, G. Tosin, M. Carara, M.N. Baibich, A.A. Gomes and
J. E. Schmidt, {\it Appl. Phys. Lett.} {\bf 61}, 12 (1992); {\it J. Magn.
Magn. Mater.} {\bf 121}, 399 (1993); G. Tosin, {\it MSc. Thesis},
Universidade Federal do Rio Grande do Sul, Porto Alegre, Brazil
(1992).\par

\item{[49]} K. Kawasaki, in {\sl Phase Transitions and Critical
Phenomena}, edited by C. Domb and M. S. Green, Academic, New York (1972),
Vol. 2, p. 443, and references therein.\par
  
\item{[50]} E.A.M. van Alphen, P.A.A. van der Heijden and  W.J.M. de
Jonge, {\it J. Magn. Magn. Mater.} {\bf 140-144}, 609 (1995).\par

\item{[51]} K.G. Wilson, {\it Sci. Am.} {\bf 241}, 140 (August 1979).\par

\vfill\eject
\parG{\bf Figure Captions}\parG

\item{I.1} Schematic representation of a water drop pending from a
vertical wall. This is the initial condition adopted in [1]. After many
boundary relaxations, the equilibrium geometrical aspect of the drop is
measured, and the results [1] are in excellent agreement with real
drops.\par

\item{II.1} Stages of one drop formation using the lattice drop model. The
values of the parameters are $W=40$, $N=150$, $g=0.1$ and $F=20$. The
numbers over the figures correspond to the time counted since the previous
drop disconnection.\par

\item{II.2} Center of mass position $versus$ time. We can see two regimes.
One before the neck formation and another when the drop starts breaking.
The usual oscillations in the center of mass position are not present
since inertia is not considered in our model. The drop here is the same
shown in the figure II.1.\par

\item{II.3} Area $versus$ perimeter, for the same drop as figure
II.1.\par

\item{II.4} Perimeter over area $versus$ time, for the same drop
again.\par

\item{II.5} First return maps of some time series obtained from the
lattice drop model, for different values of flux. The other parameters are
$g=0.5$, $N=150$, $W=20$.\par

\item{II.6} Mean fluctuation of the same time series of figure II.5. For
these and all other cases we tested, both experimental and simulated, the
Hurst exponent is $\alpha = 0.00$.\par

\item{III.1} A $log \times log$ plot of the correlations between the
moments of orders $3$ and $2$, with slope $\lambda_{3/2}$, and of orders
$5$ and $2$, with slope $\lambda_{5/2}$, are shown, as obtained in
simulations. Event-by-event analysis of experimental data exhibit the same
strong correlations. The simulations were performed on a system composed
of $216$ nucleons on a cubic $3D$ lattice, and data was extracted from a
total of $10,000$ realisations. The initial temperature, corresponding to
the excitation energy of the intermediate nucleus, was $8.50$ in units of
the Onsager critical temperature for the $2D$ Ising model. A Fermi-gas
caloric curve was used, and the surface/bulk time ratio was $a=0.998$. The
slopes satisfy the consistency conditions and yield a value for the Fisher
exponent $\tau=2.18$, consistent with $3D$ percolation. It is worth
mentioning that this value can also be obtained with a linear caloric
curve, and appears to be somewhat insensible to the value of $a$ as
well.\par

\item{III.2} The $log \times log$ plot shows the matching of the slopes of
the second moments of the fragment distributions, computed below and above
the transition, in a procedure described in the text. Data was acquired
from the same runs that were used for figure III.1. The resulting value
for $\gamma$ is consistent with the $3D$ percolation universality
class.\par

\item{IV.1} Experimental size probability distribution of the
fragments produced by a mercury drop falling onto the floor.\par

\item{IV.2} Simulated size probability distribution of fragments,
according to our dynamic lattice model. The mercury drop is represented by
a cluster of black pixels (spin $up$) on a square lattice, initially with
321 pixels. The energy is $E = 1800$, $r = 50$ and $k = 0.2$ (see text).
The plot corresponds to 5000 independent falls.\par

\item{IV.3} The same as the previous, for a cubic lattice where the
initial drop occupies 179 pixels. $E = 179$, $r = 50$ and $k = 0.9$.\par

\item{V.1} Initial magnetic grain with two domains of Ising spins, no 
external applied field.\par

\item{V.2} Snapshot of the same grain, after gradually increasing the
applied field $h$. The resulting magnetisation $m$ is also shown.\par

\item{V.3} Another snapshot. The magnetic domain wall is bent, but
not yet sliding against the grain boundaries.\par

\item{V.4} Domain wall starts to slide.\par

\item{V.5} Due to sliding, the magnetisation increases now very fast.\par

\item{V.6} Saturated magnetic grain.\par

\item{V.7} Complete hysteresis curve. It starts from the virgin grain
(figure V.1), following branch 0 as the external field gradually
increases. After saturation, the field is gradually decreased and then
inverted, the system following branch 1. Branch 2 corresponds to the
reverse path, relative to 1.\par

\item{VI.1} Schematic representation of a stripe of material A sandwiched
by B atoms. We consider a square lattice and first-neighbour
interactions.\par

\item{VI.2} Mean roughness variation as a function of annealing
temperature $T_a$ for different values of $W = $ 6 (circles), 8 (squares),
10 (diamonds), 12 (triangles), and 15 (crosses). Averages were calculated
with $N = 100$ different samples, and typical error bars are shown. Lines
are cubic splines used as guides to the eyes. The insets show snapshots of
the final $W = 6$ interface configuration at $T_0$, for some values of
$T_a$. For $T_a = 2.52$ the A layer breaks into clusters.\par

\bye